\documentstyle[12pt]{article}
\begin{document}
\def\beq{\begin{equation}}
\def\eeq{\end{equation}}
\def\bea{\begin{eqnarray}}
\def\eea{\end{eqnarray}}
\def\ve{\vert}
\def\vel{\left|}
\def\ver{\right|}
\def\nnb{\nonumber}
\def\ga{\left(}
\def\dr{\right)}
\def\aga{\left\{}
\def\adr{\right\}}
\def\rar{\rightarrow}
\def\nnb{\nonumber}
\def\la{\langle}
\def\ra{\rangle}
\def\ba{\begin{array}}
\def\ea{\end{array}}
\def\tep{$B \rar K \ell^+ \ell^-$}
\def\tepm{$B \rar K \mu^+ \mu^-$}
\def\tept{$B \rar K \tau^+ \tau^-$}
\def\ds{\displaystyle}



\def\bos{\lower 0.5cm\hbox{{\vrule width 0pt height 1.2cm}}}
\def\boss{\lower 0.35cm\hbox{{\vrule width 0pt height 1.cm}}}
\def\aaa{\lower 0.cm\hbox{{\vrule width 0pt height .7cm}}}
\def\dol{\lower 0.4cm\hbox{{\vrule width 0pt height .5cm}}}


\title{ {\Large {\bf Semileptonic $B \rar a_1 \ell \nu$ decay in QCD} } }

\author{\vspace{1cm}\\
{\small T. M. Aliev \thanks
{e-mail: taliev@rorqual.cc.metu.edu.tr}\,\,,
M. Savc{\i} \thanks
{e-mail: savci@rorqual.cc.metu.edu.tr}} \\
{\small Physics Department, Middle East Technical University} \\
{\small 06531 Ankara, Turkey} }
\date{}

\begin{titlepage}
\maketitle
\thispagestyle{empty}

\begin{abstract}
\baselineskip  0.7cm
The form factors and the branching ratio of the
$B \rar a_1 \ell \nu$ decay are calculated in framework of QCD 
sum rules. A comparison of  our results on form factors and branching 
ratio with the results from constituent quark model is presented. 

\end{abstract}

\vspace{1cm}
\end{titlepage}

\section{Introduction}
Inclusive and exclusive decays of heavy flavors play a complementary role
in determination of fundamental parameters of the Standard Model
and of a deeper understanding the dynamics QCD. Among of all these decays
the semileptonic decays occupy a special place, since their theoretical
description is relatively simple and due to the possibility of a more precise 
determination of the Cabibbo--Kobayashi--Maskawa (CKM) matrix elements
experimentally.

The new experimental results in studying $B$ meson decays, which have been
carried in recent years, has improved the values of the CKM matrix elements
and CP violation parameter. The experimental prospects at future
$B$ factories Belle \cite{R1} and BaBar \cite{R2}, where many inclusive and
exclusive channels are expected to be measured more precisely, pushes to
make further analysis on the theoretical side. One of the main goals in
these investigations is a more accurate determination of the CKM matrix
element $V_{ub}$. For this aim, both exclusive and inclusive 
$b \rar u$ transitions will be analyzed.
Studying inclusive decays theoretically is easy, but their measurement in
experiments are difficult. Using inclusive decays in determining
$V_{ub}$, implies the need of using perturbative QCD methods in the region
near the end--point of the lepton spectrum, where many resonances are
present and perturbative results are less reliable. This problem can be
avoided by considering exclusive channels for which measurements in 
experiments are easy. But unfortunately theoretical description of the
exclusive decays is not as easy as the inclusive decays. 
This is due to the fact that, in investigating
the exclusive decays there appears the problem of calculating the
hadronic  matrix elements, which is directly related to the
non--perturbative sector of QCD. So in an investigation of the exclusive
decays, we need non--perturbative methods, such as QCD sum rules, lattice
calculations, etc.

The relevant semileptonic exclusive decays in determining $V_{ub}$ are
$B \rar \pi \ell \nu$, $B \rar \rho \ell \nu$ and $B \rar a_1 \ell \nu$.
The decays $B \rar \pi \ell \nu$ and $B \rar \rho \ell \nu$, have been
extensively studied theoretically in a series of papers, in framework 
of non--perturbative approaches such as QCD sum rules \cite{R3}--\cite{R7}, 
light cone QCD sum rules \cite{R8}--\cite{R11}, lattice calculations 
\cite{R12}--\cite{R13}, quark model \cite{R14}. Note that the 
$B \rar \pi \ell \nu$ and $B \rar \rho \ell \nu$ decays have already been
observed in experiments conducted by the CLEO Collaboration \cite{R15}. 
The semileptonic decay 
$B \rar a_1 \ell \nu$ was investigated in framework of the constituent quark
model (CQM) in \cite{R16}. Yet, this decay has not been observed in
experiments that have been conducted so far, but expected to be detected in
future $B$--factories. For these reasons, it is the right time to
investigate this decay in framework of different methods and a comparison of
the predictions of different approaches is necessary. In this work we
investigate the semileptonic decay $B \rar a_1 \ell \nu$ in framework of
the three point QCD sum rules method.

The present paper is organized
as follows. In section 2 we formulate the sum rules for $B \rar a_1$
transition form factors. Section 3 is devoted to the numerical analysis of
the sum rules, and a comparison of our results with the quark model
predictions.

\section{Sum rules for $B \rar a_1$ transition form factors}

In calculation of the $B \rar a_1$ transition form factors, we start by
considering the following correlator
\bea
\Pi_{\mu\nu} (p,p^\prime,q) = i^2 \int d^4 x d^4 y e^{i p^\prime x - i p y}  
\left< 0 \vel \mbox{\rm T} \left\{ \bar d (x) \gamma_\nu \gamma_5 u(x) 
J_\mu (0) \bar b (x) \gamma_5 d (y)\right\} \ver \right>~,
\eea
where $J_\mu = \bar u \gamma_\mu (1-\gamma_5) b$ is the weak current,
with momentum $q$, 
$\bar d \gamma_\nu \gamma_5 u$ and $\bar b i \gamma_5 d$ are the currents
with quantum numbers of the final $a_1$ and initial $B$ mesons with
four--momentum $p^\prime$ and $p$, respectively. 

The correlator is expressed through the invariant amplitudes as follows:
\bea 
\Pi_{\mu\nu} = i g_{\mu\nu} \Pi_0 - i ( p + p^\prime)_\mu p_\nu \Pi_+ -
i q_\mu p_\nu \Pi_- - \varepsilon_{\mu\nu\alpha\beta} p^\alpha p^{\prime
\beta} \Pi_V + \cdots
\eea
In what follows we calculate the correlator (1) at negative values of $p^2$
and $p^{\prime 2}$ with the help of the operator product expansion in QCD,
on one side, and saturating (1) by the lowest meson states in the
pseudoscalar and axial vector channels, on the other side. Both
representations are equated using Borel transformations in $p^2$ and
$p^{\prime 2}$, which suppress higher resonance and continuum
contributions, as well as higher dimension operator contributions.

Firstly let us consider the physical part of (1). Saturating (1) by the
contributions coming from $B$ and $a_1$ mesons, we have
\bea
\Pi_{\mu\nu} (p,p^\prime) &=& \frac{1}{p^2-m_B^2} \, 
\frac{1}{p^{\prime 2}-m_{a_1}^2} \left< 0 \vel \bar d \gamma_\nu \gamma_5 u
\ver a_1 \right>
\left< a_1 \vel \bar u \gamma_\mu (1-\gamma_5) b \ver B \right>
\left< B \vel \bar b i \gamma_5 u \ver 0 \right>~.
\eea
The weak matrix element $B \rar a_1$ can be written as $(q=p-p^\prime)$ 
\bea
\lefteqn{
\left< a_1 (\varepsilon,p^\prime) \vel \bar u \gamma_\mu (1-\gamma_5) b 
\ver B(p) \right> =}\nnb \\
&&\frac{2 A (q^2)}{m_B+m_{a_1}}\, \epsilon_{\mu\nu\alpha\beta} 
\varepsilon^{\ast\nu} p^\alpha p^{\prime\beta} - 
i \varepsilon^\ast_\mu (m_B+m_{a_1}) V_1(q^2) + i (\varepsilon^\ast p)
\frac{(p+p^\prime)_\mu}{m_B+m_{a_1}} V_2(q^2) \nnb \\
&& + i (\varepsilon^\ast p) \frac{2 m_{a_1} q_\mu}{q^2}
\left[V_3(q^2) - V_0 (q^2) \right] ~,
\eea
where $m_{a_1}$ and $\varepsilon$ are the mass and the four--polarization
vector of the $a_1$ meson. Note that form factor $V_3$ can be related to 
$V_1$ and $V_2$ in the following way
\bea
V_3(q^2) = \frac{m_B + m_{a_1}}{2 m_{a_1}} V_1(q^2) -
\frac{m_B - m_{a_1}}{2 m_{a_1}} V_2(q^2)~,
\eea
and with initial condition 
\bea
V_3(0) = V_0(0) ~. \nnb
\eea
The vacuum--to--meson transition matrix elements are are defined in standard
way, namely
\bea
\left< B \vel \bar b i \gamma_5 d \ver 0 \right> &=& f_B \frac{m_B^2}{m_b}
~, \nnb \\
\left< 0 \vel \bar d \gamma_\nu \gamma_5 u \ver a_1 \right> &=& 
\frac{\sqrt{2} m_{a_1}^2}{g_{a_1}} \varepsilon_\nu~.
\eea
Secondly, let us now consider the theoretical part of the sum rules. The
theoretical part of the sum rules is calculated by means of the operator
product expansion for the correlator (1). Up to dimension 6, the operators
are determined by the contribution of the bare loop, and power corrections
coming from dimension--3 $\la \bar \psi \psi \ra$, dimension--4 
$\la G^2 \ra$ dimension--5 $m_0^2 \la \bar \psi \psi \ra$, and dimension--6
$\la \bar \psi \psi \ra^2$ operators. Our calculations show that the contributions
coming from $\left< G^2 \right>$ and $\la \bar \psi \psi \ra^2$ are 
negligibly small, and for this reason we
don't take their contribution into consideration in the present analysis.

The double dispersion relation, for the perturbative contribution, can be
written as follows
\bea
\Pi_i = \int \!\!\!\int ds\, ds^\prime \frac{\rho_i(s,s^\prime,q^2)}
{(s-p^2) (s^\prime - p^{\prime 2})} + \mbox{\rm sub. terms}~,
\eea
where index $i$ describes the necessary invariant amplitudes 
(($i = 0;~+;~-;~V$) (see Eq. (2)), and $\rho_i$ is the corresponding
spectral density. The spectral density is obtained from the usual Feynman
integral for the bare loop by replacing 
\bea
\frac{1}{p^2} \rar - 2 \pi i  \delta (p^2)~. \nnb
\eea
After standard calculations for the spectral densities, we have
\bea
\rho_V &=& \frac{3 m_b}{4 \pi^2} \, \frac{s^\prime (s-2 m_b^2 - s^\prime +q^2)}
{\lambda^{3/2}}~, \nnb \\ \nnb \\
\rho_0 &=& - \frac{3 m_b}{8 \pi^2} \, \left[ \frac{s^\prime}{\lambda^{1/2}}
+ \frac{2 s^\prime}{\lambda^{3/2}} \ga m_b^4 + s q^2 - m_b^2 (s+q^2-s^\prime) \dr
\right] ~, \nnb \\ \nnb \\
\rho_\pm &=& \frac{3 m_b}{8 \pi^2} \, \Bigg\{ - \frac{1}{\lambda^{3/2}}
s^\prime ( s^\prime - q^2 - s + 2 m_b^2 ) + 
\frac{2 s^\prime}{\lambda^{5/2}}\Bigg[ s^\prime \ga 6 m_b^4 + \lambda
+ 6 s q^2 - 6 m_b^2 (s + q^2 - s^\prime) \dr \nnb \\
&& \mp 3 m_b^4 (s + s^\prime - q^2) \mp \lambda (s + 2 m_b^2) \mp
6 m_b^2 s (s^\prime + q^2 - s) \mp 3 s q^2 (s + s^\prime - q^2) \Bigg]
\Bigg\}~,
\eea
where $\lambda = s^2 + s^{\prime 2} + q^4 - 2 s q^2 - 2 s^\prime q^2 - 
2 s s^\prime$ is the usual triangle function. For the power correction
contributions we get 
\bea 
\Pi_V^{(3)} &=& \frac{\left< \bar \psi \psi \right>}
{(p^2-m_b^2) p^{\prime 2}}~, \nnb \\ \nnb \\
\Pi_0^{(3)} &=& \frac{(m_b^2-q^2)\left< \bar \psi \psi \right>}
{2 (p^2-m_b^2) p^{\prime 2}}~, \nnb \\ \nnb \\
\Pi_\pm^{(3)} &=& \pm \, \frac{\left< \bar \psi \psi \right>}
{2 (p^2-m_b^2) p^{\prime 2}}~, \nnb \\ \nnb \\
\Pi_V^{(5)} &=& - m_0^2 \left<\bar \psi \psi \right> 
\Bigg[ \frac{1}{3 (p^2-m_b^2)^2 p^{\prime 2}}
+ \frac{m_b^2}{2(p^2-m_b^2)^3 p^{\prime 2}} +
\frac{m_b^2-q^2}{3 (p^2-m_b^2)^2 (p^{\prime 2})^2} \Bigg]~, \nnb \\ \nnb \\
\Pi_0^{(5)} &=&  - m_0^2 \left<\bar \psi \psi \right> \Bigg[ - \frac{1}{6 (p^2-m_b^2) p^{\prime 2}} +
\frac{(m_b^2-q^2)^2}{6(p^2-m_b^2)^2 (p^{\prime 2})^2}+
\frac{m_b^2-q^2}{6(p^2-m_b^2) (p^{\prime 2})^2}  \nnb \\ \nnb \
&+& \frac{3 m_b^2-4 q^2}{12 (p^2-m_b^2)^2 p^{\prime 2}} +
\frac{m_b^4 - m_b^2 q^2}{4 (p^2-m_b^2)^3 p^{\prime 2}} \Bigg] ~, 
\nnb \\ \nnb \\
\Pi_+^{(5)} &=& m_0^2 \left<\bar \psi \psi \right> \Bigg[
\frac{1}{6 (p^2-m_b^2)^2 p^{\prime 2}} -
\frac{m_b^2}{4 (p^2-m_b^2)^3 p^{\prime 2}} -
\frac{m_b^2-q^2}{6 (p^2-m_b^2)^2 (p^{\prime 2})^2}\Bigg] ~,\nnb \\ \nnb \\
\Pi_-^{(5)} &=& m_0^2 \left<\bar \psi \psi \right>
\Bigg[\frac{1}{2 (p^2-m_b^2)^2 p^{\prime 2}} +
\frac{m_b^2}{4 (p^2-m_b^2)^3 p^{\prime 2}} + 
\frac{m_b^2-q^2}{6 (p^2-m_b^2)^2 (p^{\prime 2})^2}\Bigg]~.
\eea
In Eqs. (8) and (9) subscripts within the parentheses denote the dimension of
the corresponding operators. Using Eqs. (3), (4), (6), (7), (8) and (9) and
using double Borel transformation in variables $p^2$ and $p^{\prime 2}$ and
equating two different representations of the correlator (1) we get the
following sum rules for the form factors describing $B \rar a_1$ transition:
\bea
A(q^2) &=& \frac{m_b g_{a_1} (m_B+m_{a_1})}{2 \sqrt{2} f_B m_B^2 m_{a_1}^2} \,
e^{(m_B^2/M_1^2 + m_{a_1}^2/M_2^2)} \nnb \\
&\times& \Bigg\{ \int\!\!\!\int ds \, ds^\prime \rho_V (s,s^\prime,q^2) 
e^{-s/M_1^2 - s^\prime/M_2^2} + \left< \bar \psi \psi \right> 
e^{-m_b^2/M_1^2} \nnb \\
&-& m_0^2 \left< \bar \psi \psi \right> e^{-m_b^2/M_1^2}
\Bigg[ - \frac{1}{3 M_1^2} + \frac{m_b^2}{4 M_1^4} 
+ \frac{m_b^2-q^2}{3 M_1^2 M_2^2} \Bigg] \Bigg\}~, \nnb \\ \nnb \\ \nnb \\
V_1(q^2) &=& \frac{m_b g_{a_1}}{\sqrt{2} f_B m_B^2 m_{a_1}^2 (m_B+m_{a_1})}
e^{(m_B^2/M_1^2 + m_{a_1}^2/M_2^2)} \nnb \\
&\times& \Bigg\{ \int\!\!\!\int ds \, ds^\prime \rho_0 (s,s^\prime,q^2)
e^{-s/M_1^2 - s^\prime/M_2^2} + \left< \bar \psi \psi \right>
\frac{m_b^2-q^2}{2} e^{- m_b^2/M_1^2} \nnb \\
&+& m_0^2 \left< \bar \psi \psi \right> e^{-m_b^2/M_1^2}
\Bigg[\frac{1}{6} - \frac{(m_b^2-q^2)^2}{6 M_1^2 M_2^2}+
\frac{m_b^2-q^2}{6 M_2^2} \nnb \\
&+&\frac{3 m_b^2-4 q^2}{12 M_1^2} 
- \frac{m_b^4-m_b^2 q^2}{8 M_1^4} \Bigg] \Bigg\}~,  \nnb \\ \nnb \\ \nnb \\
V_2(q^2) &=& \frac{m_b g_{a_1} (m_B+m_{a_1})}{\sqrt{2} f_B m_B^2 m_{a_1}^2}
e^{(m_B^2/M_1^2 + m_{a_1}^2/M_2^2)} \nnb \\
&\times& \Bigg\{ \int\!\!\!\int ds \, ds^\prime \rho_+ (s,s^\prime,q^2)
e^{-s/M_1^2 - s^\prime/M_2^2} + \frac{1}{2} \left< \bar \psi \psi \right>
e^{- m_b^2/M_1^2} \nnb \\
&-& m_0^2 \left< \bar \psi \psi \right> e^{-m_b^2/M_1^2}
\Bigg[\frac{1}{6 M_1^2} +\frac{m_b^2}{8 M_1^4}+
\frac{m_b^2-q^2}{6 M_1^2 M_2^2}  \Bigg] \Bigg\}~, \nnb \\ \nnb \\ \nnb \\
V_3(q^2)-V_0(q^2) &=& \frac{m_b g_{a_1} q^2}{2 \sqrt{2} f_B m_B^2 m_{a_1}^3}\,
e^{(m_B^2/M_1^2 + m_{a_1}^2/M_2^2)} \nnb \\
&\times& \Bigg\{ \int\!\!\!\int ds \, ds^\prime \rho_-(s,s^\prime,q^2)
e^{-s/M_1^2 - s^\prime/M_2^2} - \frac{1}{2} \left< \bar \psi \psi \right>
e^{-m_b^2/M_1^2} \nnb \\
&+& m_0^2 \left< \bar \psi \psi \right> e^{-m_b^2/M_1^2}
\Bigg[ - \frac{1}{2 M_1^2} + \frac{m_b^2}{8 M_1^4} 
+ \frac{m_b^2-q^2}{6 M_1^2 M_2^2} \Bigg] \Bigg\}~.
\eea
The region of integration, which appears in calculation of the perturbative
contribution, is determined by the following inequalities:
\bea
0<&s^\prime&< s^\prime_0 ~, \nnb \\
m_b^2 + \frac{m_b^2}{m_b^2-q^2} s^\prime<&s&<s_0~, 
\eea
where $s_0$ and $s^\prime_0$ are the continuum thresholds in the $B$ and
$a_1$ meson channels, respectively. In Eq. (10), the continuum contribution
is modeled as the bare loop contribution starting from the thresholds
$s_0$ and $s^\prime_0$ and subtracted from bare loop contribution.

Finally, using the expressions for the form factors, we present the
differential decay width $d\Gamma/dq^2$ for the $B \rar a_1 \ell \nu$, which
can be written in terms of the helicity amplitudes
\bea
H_\pm(q^2) &=& (m_B+m_{a_1}) V_1(q^2) \mp \frac{\lambda^{1/2}(m_B^2,m_{a_1}^2,q^2)}
{m_B+m_{a_1}} A(q^2) ~, \nnb \\ \nnb \\
H_0(q^2) &=& \frac{1}{2 m_{a_1} \sqrt{q^2}} \Bigg[
(m_B^2-m_{a_1}^2-q^2) (m_B+m_{a_1}) V_1(q^2) - 
\frac{\lambda(m_B^2,m_{a_1}^2,q^2)}{m_B+m_{a_1}} V_2(q^2) \Bigg]~, \nnb 
\eea
in the following way
\bea
\frac{d \Gamma_\pm}{d q^2} &=& \frac{G^2 \vel V_{ub} \ver^2}{192 \pi^3 m_B^3}
q^2 \lambda^{1/2}(m_B^2,m_{a_1}^2,q^2) \vel H_\pm \ver^2 ~, \nnb \\ \nnb \\
\frac{d \Gamma_0}{d q^2} &=& \frac{G^2 \vel V_{ub} \ver^2}{192 \pi^3 m_B^3}
q^2 \lambda^{1/2}(m_B^2,m_{a_1}^2,q^2) \vel H_0 \ver^2 ~,
\eea
where $\pm,~0$ refers to the $a_1$ helicities.
Note that the difference of the form factors $V_3-V_0$ does not give any
contribution to the $B \rar a_1 \ell \nu$ decay, since it is
proportional to the lepton mass (in our case electron or muon).
\section{Numerical analysis}
In calculation of the form factors $A(q^2),~V_1(q^2),~V_2(q^2)$ and $V_0(q^2)$, 
we use the following values of the input parameters: $m_b=4.8~GeV$, 
$m_{a_1}=1.26~GeV$, $g_{a_1}=8.9$ \cite{R17}, $m_B = 5.28~GeV$,
$s_0^\prime=3 ~GeV^2$, 
$\left< \bar \psi \psi \right>= - (0.24~GeV)^3$ \cite{R18}, 
$m_0^2 = (0.8 \pm 0.2)~GeV^2$
\cite{R19} (at the normalization point $\mu=1~GeV$). 
For the value of the leptonic decay
constant, we shall use $f_B=140~GeV$. This value of $f_B$ corresponds to the
case where ${\cal O}(\alpha_s)$ corrections are not taken into account (see
\cite{R20,R21}). Since the bare loop contribution does not
involve ${\cal O}(\alpha_s)$  corrections, they are not taken into
consideration in calculation of the leptonic decay constant as well, for
sake of a consistent procedure. 

The expressions for the form factors involves two independent parameters
$M_1^2$ and $M_2^2$. According to the QCD sum rules ideology, the problem is
to find a region where the results are practically independent of $M_1^2$
and $M_2^2$, while at the same time, the power corrections and continuum 
contributions remain under control.
On evidence of the existing calculations (see \cite{R22}), we expect the
region of stability for three--point correlator to be at values of Borel
parameters twice as large as in the corresponding two--point functions. From
an analysis of the two--point functions, it follows that the stability
region is $4~GeV^2 < M_1^2 < 8~GeV^2$ ((see \cite{R20}). Therefore we find
it convenient to evaluate in the range $8~GeV^2 < M_1^2 < 15~GeV^2$. In
determining the range of the other independent parameter $M_2^2$, we have
used the following relation,
\bea
\frac{M_2^2}{M_1^2} \simeq \frac{m_{a_1}^2}{m_B^2-m_b^2} \simeq
\frac{1}{3}~,
\eea
from which it follows that $M_2^2 \simeq M_1^2/3$. We have checked our
results weak dependence on this ratio.

In Figs. 1 (a)--(d), we present the $M_1^2$ dependence of the form factors
$A(q^2=0),~V_1(q^2=0),~V_2(q^2=0)$ and $V_0(q^2=0)$ at fixed values 
$M_2^2/M_1^2 = 1/3$, 
$s_0^\prime=3~GeV^2$ and at three different values of the threshold, 
$s_0=33~GeV^2,~s_0=35~GeV^2,~s_0=37~GeV^2~$. From these figures we 
observe that the variations in the results, with changing $s_0$, are 
about $\sim 5\%$.
It should be noted here that, in all cases of interest, the main contribution
comes from the dimension 3 operator $\left< \bar \psi \psi \right>$.   
Our final results for the values of the form factors at $q^2=0$ are
presented in Table 1, in which we present the CQM
predictions at $q^2=0$ as well. A comparison of these results yields that
the magnitude of the form factor  $A(q^2=0)$ is approximately five times larger,
while the magnitude of the form factors $V_0(q^2=0)$ and $V_1(q^2=0)$ are
five and two times smaller, respectively,  in our case, than the ones 
predicted by the CQM (the errors in form factors predicted by CQM is
about $15\%$, see \cite{R16}). The magnitude of the form factor 
$V_2(q^2=0)$ is practically the same in both approaches.   
Figs. 2 (a)--(d), depict the $q^2$ dependence of the form factors at fixed values
$M_1^2=10~GeV^2$ and $M_1^2=12~GeV^2$, under the condition 
$M_2^2/M_1^2 = 1/3$, and at
$s_0=35~GeV^2$, in the range $0<q^2<10~GeV^2$. The reason why we consider 
this region of $q^2$ is that the non--perturbative contribution becomes large
and hence the operator product expansion breaks down. In order to extend our
results to the full physical region 
$q_{max}^2 \simeq 16~GeV^2$, we have used the
following extrapolation formula for the form factors in such a way that
they reproduce the sum rules prediction up to $10~GeV^2$
region. We have found that the best fits, which satisfy the above--mentioned
condition, can be expressed as
\bea
F_i(q^2) = \ds{\frac{F_i(0)}
{1 - a_F\ga q^2/m_B^2 \dr + b_F \ga q^2/m_B^2 \dr^2 }}~. \nnb 
\eea
The values of the constant fit parameters $a_F$ and $b_F$, for different
form factors, are listed in Table 1. 
After integrating Eq. (12) over $q^2$, and using $V_{ub}=0.032$, $\tau_B =
1.56\times 10^{-12}~s$, one can obtain the decay width and
branching ratio of the $B \rar a_1 \ell \nu$ process.
These results, together with CQM predictions, are summarized in Table 2, 
which constitute the
main results of this work. 
From a comparison of these results we see that, the CQM's
prediction of branching ratio is approximately five times larger than
the QCD sum rules prediction. 
\begin{table}
\begin{center}
\begin{tabular}{|c|ccc||c|}
\hline
                   &\boss This work & $a_F$ & $b_F$ & 
CQM \cite{R16} \\ \hline \hline
$~A(0)~$ &\boss $ -0.67 \pm 0.10$ & $\phantom{-}0.72$ &$-0.20$ &
0.14\\ \hline  
$~V_1(0)~$ &\boss $-0.42 \pm 0.05$ & $-0.44$ &$\phantom{-}0.45$ &
 $0.81$\\ \hline
$~V_2(0)~$ & \boss $- 0.53 \pm 0.05$ & $\phantom{-}0.45$ & $\phantom{-}0.13$ &
 $0.56$\\ \hline
$~V_0(0)~$ &\boss $- 0.23 \pm 0.05$ & $\phantom{-}0.86$ & $- 0.38$ &
$1.20$ 
\\ \hline   
\end{tabular}
\vskip 0.3 cm
\caption{}  
\end{center}
\end{table} 
\begin{table}
\begin{center}
\begin{tabular}{|c|c|c|}
\hline
                   &\boss This work & CQM \cite{R16}       \\ \hline \hline
\boss $\Gamma_+(B \rar a_1 \ell \nu)$ & $5.1 \times 10^5~s^{-1}$ & 
$(4.6 \pm 0.9) \times 10^7~s^{-1}$  \\ \hline
\boss $\Gamma_-(B \rar a_1 \ell \nu)$ & $7.5 \times 10^7~s^{-1}$ & 
$(0.98 \pm 0.18)\times 10^8~s^{-1}$ \\ \hline
\boss $\Gamma_0(B \rar a_1 \ell \nu)$ & $2.7 \times 10^7~s^{-1}$ & 
$(4.0 \pm 0.7)\times 10^8~s^{-1}$   \\ \hline
\boss ${\cal B}(B \rar a_1 \ell \nu)$ & $1.6 \times 10^{-4}$ & 
$(8.4 \pm 1.6)\times 10^{-4}$ 
\\ \hline
\end{tabular}
\vskip 0.3 cm
\caption{}
\vskip 1 cm
\end{center}
\end{table}

\newpage
\section*{Figure captions}
{\bf Fig. 1} The dependence of the form factors $A(q^2=0),~
V_1(q^2=0),~V_2(q^2=0),$ and $V_0(q^2=0)$ on Borel parameter $M_1^2$. 
In all figures (a)--(d), the dotted line corresponds to value of threshold 
$s_0=33~GeV^2$, the solid line corresponds to $s_0=35~GeV^2$, and
dash--dotted line corresponds to $s_0=37~GeV^2$, respectively.\\ \\   
{\bf Fig. 2} The dependence of the form factors
$A(q^2)~,V_1(q^2)~,V_2(q^2)~$ and $V_0(q^2)$ on $q^2$ at fixed value of
threshold $s_0=35~GeV^2$. In all figures (a)--(d), the solid line
corresponds to the fixed value of the Borel parameter $M_1^2=10~GeV^2$, and 
the dash--dotted line corresponds to $M_1^2=12~GeV^2$, respectively.

\newpage
\begin{figure} 
\vspace{32.0cm}
    \includegraphics{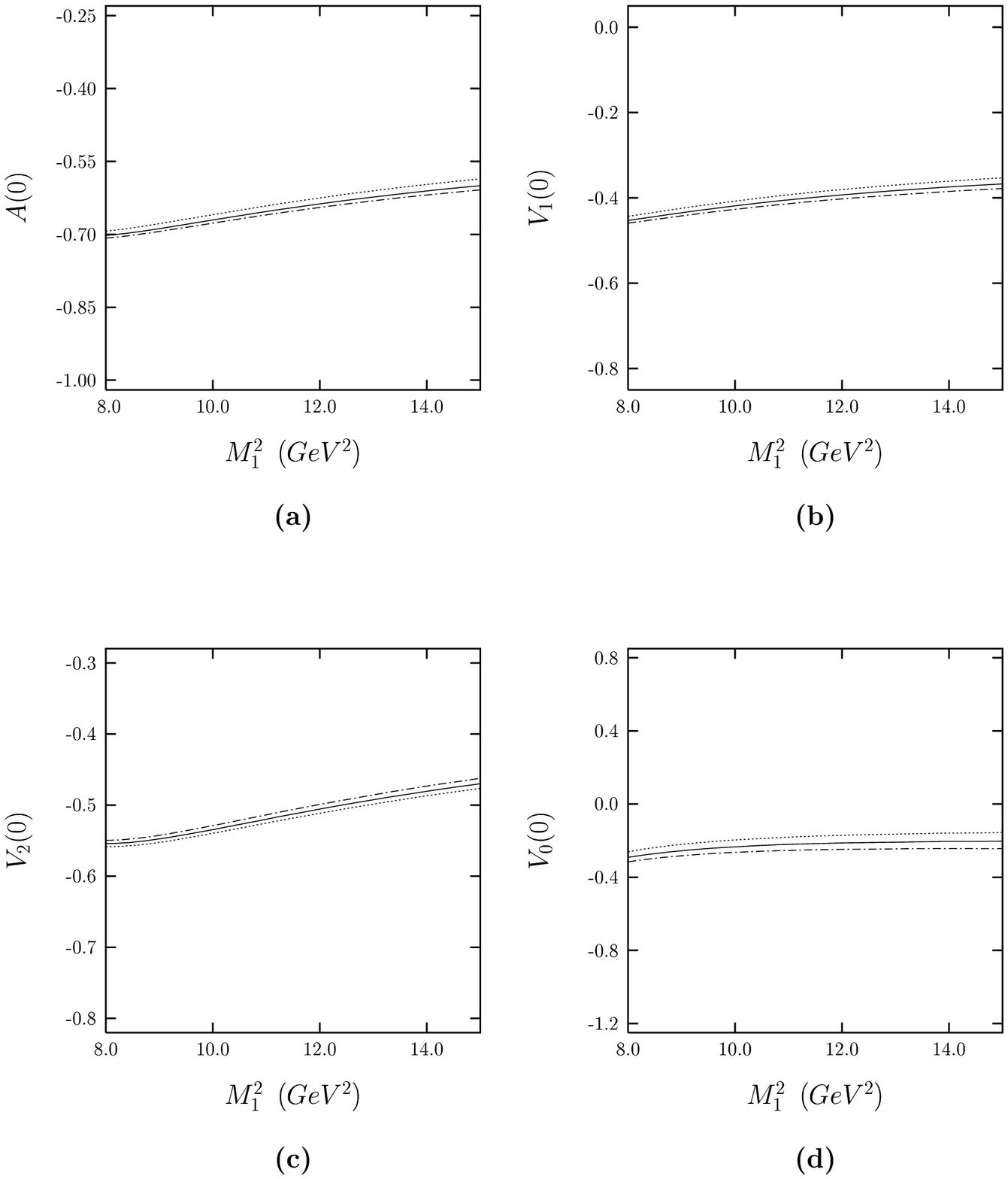}
    \vspace{-12.0cm}
\caption{}
\end{figure}
\begin{figure}
\vspace{32.0cm}
    \includegraphics{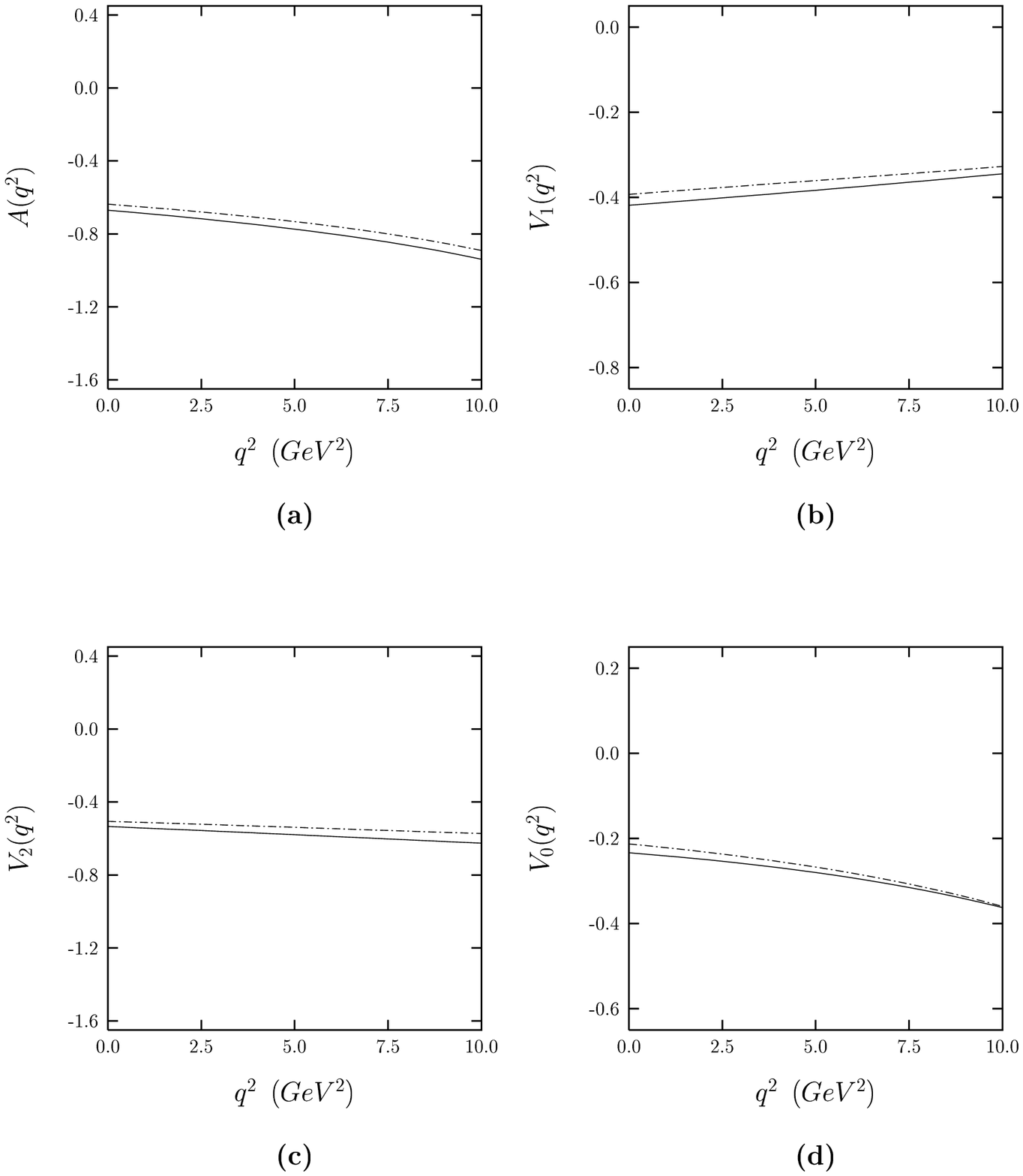}
    \vspace{-12.0cm}
\caption{}
\end{figure}

\end{document}